\documentclass[reprint,amsmath,amssymb,aps]{revtex4-1}
\usepackage{amsmath}
\usepackage{epsfig}
\usepackage{epstopdf}
\begin{document}
\title{The rhythm of coupled metronomes}
%\subtitle{}
\author{Sz. Boda, Z. N\'eda, B. Tyukodi and A. Tunyagi}

\affiliation{Babe\c{s}-Bolyai University, Department of Physics, str. Kog\u{a}lniceanu 1 nr.~1, 400084,
Cluj-Napoca, Romania}

\date{Received: date / Revised version: date}
% The correct dates will be entered by Springer
%
\begin{abstract}
Spontaneous synchronization of an ensemble of metronomes placed on a freely rotating platform is
studied experimentally and by computer simulations.  A striking in-phase synchronization is observed when the metronomes' beat frequencies are fixed above a critical limit.
Increasing the number of metronomes placed on the disk leads to an observable decrease in the level of the emerging synchronization. 
A realistic model with experimentally determined parameters is considered in order to understand the 
observed results. The conditions favoring the emergence of synchronization are investigated. It is shown that the 
experimentally observed trends can be reproduced by assuming a finite spread in the metronomes' natural frequencies. 
In the limit of large numbers of metronomes, we show that synchronization emerges only above a critical beat frequency value.    
\end{abstract}

 % end of PACS codes
 %end of abstract
%

\maketitle

\section{Introduction}

Spontaneous synchronization of coupled non-identical oscillators is a well-known form of
collective behavior\cite{book:strog,sync,book:pik}. The problem has been intensively studied since Huygens. If the legend is true, presumably he was the first one who noticed and reported the synchronized swinging of pendulum clocks. By simple ex\-pe\-ri\-ments he found that the synchronized state is a stable limit cycle of the system, 
because even after perturbing the system, the pendulums came back to this dynamic state.
Originally, Huygens thought that this \emph{"odd kind of sympathy"}, as he named it, occurs due to shared air currents between the pendulums. He performed several tests to confirm this idea.
His experimental setup was really simple, with two pendulum clocks hung from a common suspension beam which was placed between two chairs \cite{book:pik} .
After performing some additional tests, Huygens observed a stable and reproducible anti-phase synchronization, and attributed this to imperceptible vibrations in the suspension beam.
He summarized his observations in a letter to the Royal Society of London \cite{huy}, and launched the study of synchronization phenomena and coupled oscillators.

Recent studies have aimed at reconsidering various forms of Huygens' two pedulum-clock experiment as well as realistically modelling the system.  Bennet and his group \cite{bennet} investigated the same two pendulum-clock system as Huygens did and came to the conclusion that several types of collective dynamics are observable as a function of the system's parameters. For strong coupling, a "beat death" phenomenon usually occurs where one pendulum oscillates and the other does not. For weak coupling, synchronization does not occur, and a quasi-periodic oscillation is observed. There is, however, an intermediate coupling strength interval where
the anti-phase synchronization observed by Huygens appears. Hence, Huygens had some luck with his setup as the coupling was just in the right interval: strong enough to cause  synchronization, but also weak enough to avoid the "beat death" phenomenon.

Dilao \cite{dilao} came to the conclusion that the periods of two synchronized nonlinear oscillators (pendulum clocks) differ from the natural frequencies of the oscillators. Kumon and his group \cite{kumon} have studied a similar system consisting of two pendulums, one of them having an excitation mechanism, 
and the two pendulums being coupled by a weak spring. Fradkov and Andrievsky \cite{fradkov} developed a model for such a system, and obtained from numerical solutions that both in- and anti-phase synchronizations are possible, depending on the initial conditions.

Kapitaniak and his group \cite{czo} revisited Huygens' original experiment and found that the anti-phase 
synchronization usually emerges, although in rare cases in-phase synchronization is also possible. They also developed a more realistic model for this experiment \cite{kapit}. 

Pantaleone \cite{panta} considered a similar system, but he used metronomes placed on a freely moving base (suspended on two 
cylinders) instead of pendulum clocks. He modeled the metronomes as van der Pol oscillators \cite{vanderpol} and came to the conclusion that anti-phase synchronization occurs in some rare cases only. He proposed this setup as an easy classroom demonstration for the Kuramoto model \cite{kuramoto} and extended the study for larger systems containing up to seven globally coupled metronomes. He also made quantitative investigations by tracking the motion of the metronomes' pendulum by acoustically registering the ticks with a microphone. Ulrichs and his group \cite{ulric} examined the case when the number of metronomes was even larger.

The present state of this quite old field of physics was recently reviewed by Kapitaniak et. al  \cite{Kap2012}.

Our work is intended to continue this line of studies showing that it is still possible to find interesting 
aspects of this quite old problem in physics. In contrast to previous works, we consider an ensemble of metronomes 
arranged symmetrically on the perimeter of a freely rotating disk, as illustrated in Figure \ref{fig1}. The free
rotation of the disk acts as a coupling mechanism between the metronomes and, for high enough ticking frequencies, synchronization emerges. Our aim here is to investigate the conditions favoring such spontaneous synchronization by
using a realistic model and model parameters. In order to achieve our task, we first study the dynamics of the system by well-controlled experiments. Contrary to earlier studies that investigated only the final stable dynamic state of the system, here we also consider and describe the transient dynamics leading to synchronization. The synchronization level is quantified and measured. This is achieved by using an optical phase-detection mechanism for each metronome separately. We then construct a realistic model for the system and its modeling power is proved by comparing its results with the experimental ones. We discuss the reasons behind the fact that only in-phase synchronization is observed in our experiments. Finally, the model is used to investigate the  emergence of synchronization in large ensembles of coupled metronomes.

\section{Experimental setup}

\begin{figure}[h]
  \centering
  \resizebox{0.40\textwidth}{!}{%
  \includegraphics{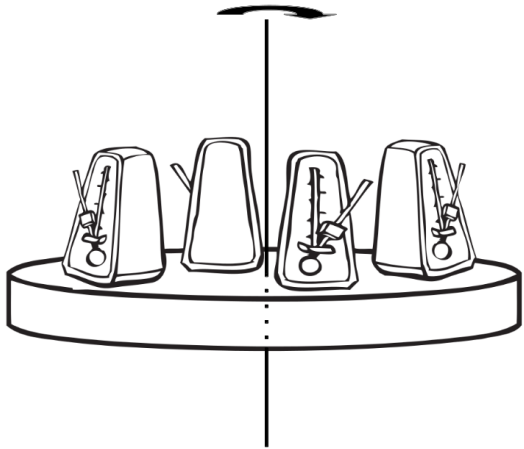}
   }
\caption{Schematic view of the experimental setup: metronomes are placed on the perimeter of
a disk that can rotate around a vertical axes.}
\label{fig1}
\end{figure}

\begin{figure}[h]
  \centering
   \resizebox{0.30\textwidth}{!}{%
   \includegraphics{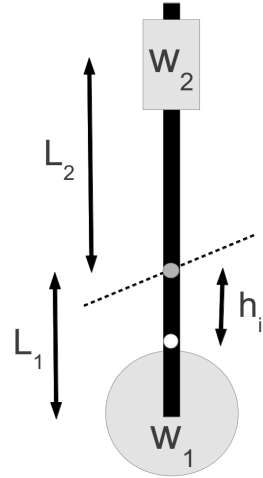}
   }
\caption{Schematic view of the metronomes' bob. The dotted line denotes the horizontal suspension axis, and the white dot illustrates the center of mass. }
\label{fig2}
\end{figure}

\begin{figure}[ht!]
  \centering
  
  \resizebox{0.40\textwidth}{!}{%
  \includegraphics{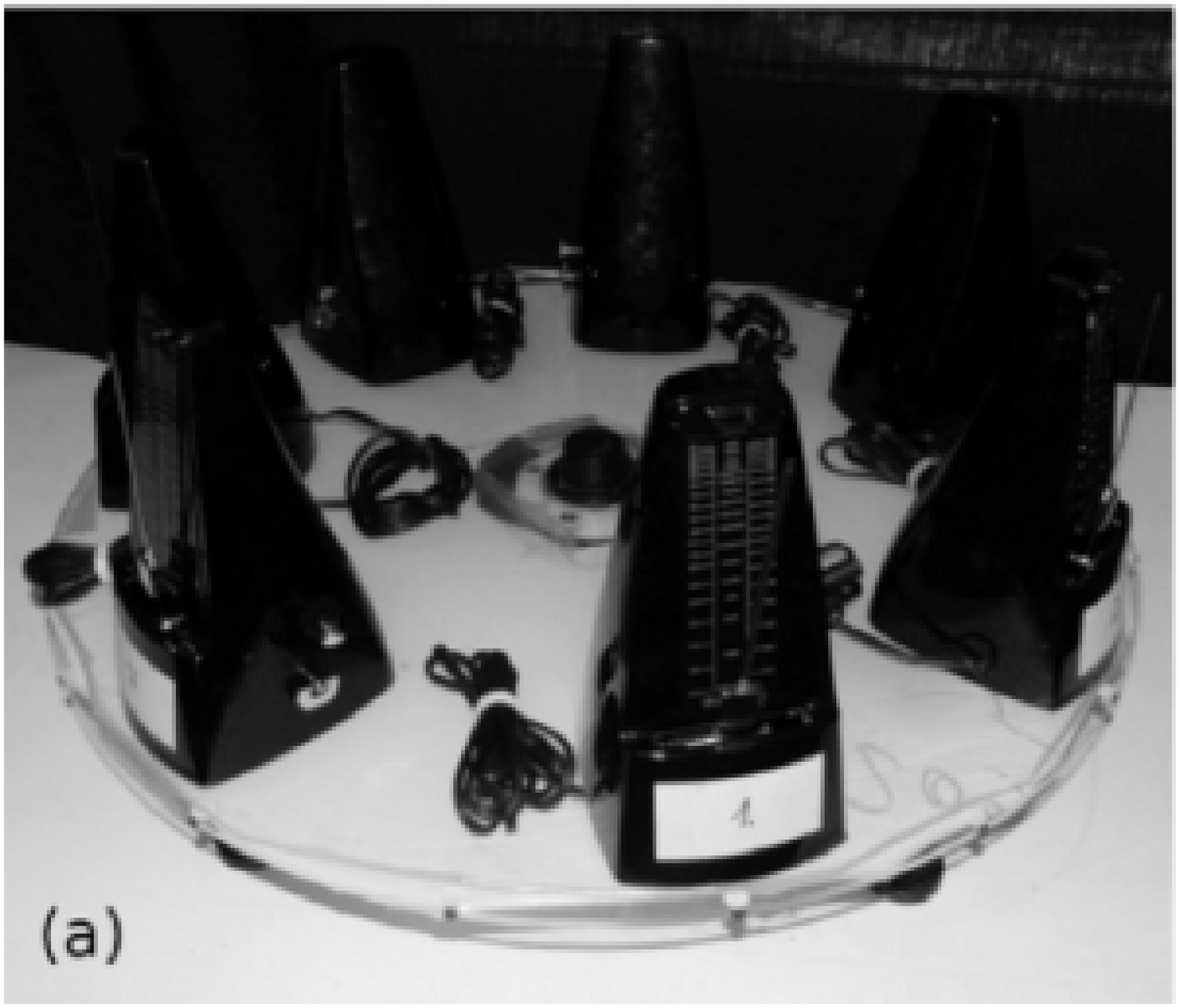}
}
 \resizebox{0.40\textwidth}{!}{%
  \includegraphics{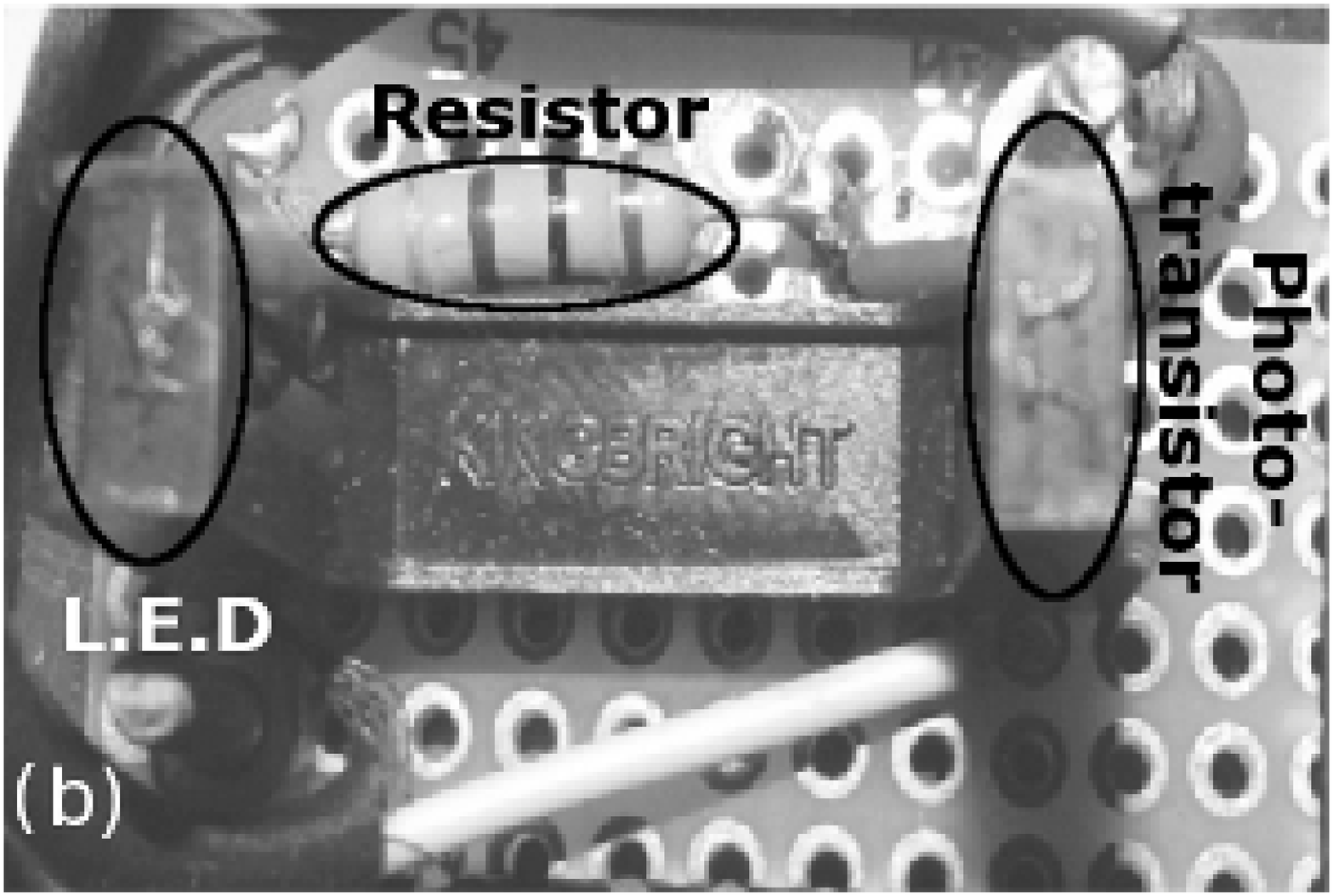}
}
   
 \caption{(a) The experimental setup, with the metronomes placed on the platform and the 
wiring that carries information on the metronomes' phases. (b) One of the light-gates (Kingbright KTIR 0611 S), composed of an infrared LED and a photo-transistor.}
\label{fig3}
 % rng.png: 640x400 pixel, 72dpi, 11x7 cm, bb=0 0 640 400
\end{figure}

The experimental setup is sketched in Figure \ref{fig1}. The main units are the metronomes (Figure 
\ref{fig1} and \ref{fig3}a), which are devices that produce regular, metrical beats. They were patented by Johann Maelzel in 1815 as a timekeeping tool for musicians (\cite{urlmet}). The oscillating element of the metronome is a physical pendulum, which consists of a rod with two weights on it (Figure \ref{fig2}):  a fixed one at the lower end of the rod, whose mass is denoted by W1, and a movable one, $W_2$, attached to the upper part of the rod. In general, $W_1>W_2$ and 
the rod is suspended on a horizontal axis between the two weights in a stable manner, so that the center 
of mass lies below the suspension axes. 

By sliding the $W_2$ weight along the rod, the oscillation frequency can be tuned. There are several marked 
places on the rod where the $W_2$ weight has a stable position, yielding standard ticking frequencies for the metronome. These $\omega_0$ 
frequencies are marked on the metronome in units of Beats Per Minute (BPM).

Another key part of the metronomes is the excitation mechanism, which compensates for the energy lost to friction. This mechanism gives additional momentum to the physical 
pendulum in the form of pulses delivered at a given phase of the oscillation period. 
For a more detailed analysis of this excitation mechanism we recommend the work  of  Kapitaniak \emph{et. al.} \cite{kapit}

For the experiments, we used the commercially available Thomann 330 metronomes (Figure \ref{fig3}a).
 From the 10 metronomes 
we had bought, the 7 with the most similar frequencies were selected.  Naturally, since there are no two identical units, we have to deal with a non-zero standard deviation of the natural frequencies in experiments

In order to globally couple the metronomes, we placed them on a disk shaped platform which could rotate with a very little friction around a vertical axis, as is sketched in Figure \ref{fig1} and illustrated in the photo in Figure \ref{fig3}a. 

  In order to monitor the dynamics of all metronomes separately, photo-cell detectors (Figure \ref{fig3}b) 
were mounted on them. 
These detectors were commercial ones (Kingbright KTIR 0611 S), and contained a Light Emitting Diode and
a photo-transistor. They were mounted on the bottom of the metronomes.

The wires starting from each metronome (seen in Figure \ref{fig3}) connect the photo-cells with a circuit board, allowing data collection through the USB port of a computer. The data was collected using a free, open-source program, called \emph{MINICOM}. (\cite{minicom}).
The data was saved in log files, and could be processed in real-time. It was possible to simultaneously follow 
 the states of up to $8$ metronomes. The circuit board only sent data 
when there was a change in the signal from the photo-cell system 
(i.e. a metronome's bob passed the light-gate). At that point, it would record a string such as $0-0-1-1-0-1-0 $ $1450$, 
where the first $7$ numbers characterize the metronome bob's position relative to the photo-cell (whether the
gate is open or closed) and the eighth number is the time, where one time unit corresponds to 64 microseconds.
Since we are interested in the dynamics of this system from the perspective of synchronization, we computed the classical order parameter, r, of the Kuramoto model \cite{kuramoto} in our numerical evaluations:
\begin{equation}
 r\exp(i\phi)=\frac{1}{N}\sum_j{\exp(i\theta_j)}.
 \label{op}
\end{equation} 
Here, $\phi$ is the average phase of the whole ensemble, $\theta_j$ is the phase of the $j$-th metronome, $N$ is the number of
metronomes,  and $i$ is the imaginary unit. 

The recorded data only tells us the exact moment at which the metronome's bob passes through the light-gate, so some additional steps are needed in order to get 
the phases $\theta_j$ of all metronomes  and to compute the Kuramoto order-parameter for a given moment in time.  In order to achieve this, we first excluded 
from the data those time-moments when the metronome's bob passed through the light-gate for the second time in a period, and after that we retained the pass-times 
corresponding to a given directional motion only. With this
"cleaned data", we calculated the period of each cycle and interpolated this time-interval 
for the $\theta_j$ phases (between $0$ and $2 \pi$, corresponding to the state of a Kuramoto rotator)
assuming a uniform angular-velocity.
This way, the phase $\theta_j$ of each metronome (considered here as a rotator) could be uniquely determined at each moment in time, and the Kuramoto order parameter (\ref{op}) could be computed.

Before starting the experiments we monitored each metronome separately and recorded their exact
frequency, $\omega_i$, for all the standardly marked rhythms. 
These frequencies had a small, but finite fluctuation around
the nominal frequency, $\omega_0$. We have selected those $7$ met\-ro\-no\-mes 
that had their $\omega_i$ standard frequencies relatively close to each other, and precisely measured these values.
From these values the standard deviation, $\sigma$, of the used metronomes' natural frequencies could be determined (Table \ref{table2}). 

\section{Experimental Results}

As already described in the introductory section, the met- metronomes oscillate with different natural frequencies, 
depending on the position of the adjustable weight on the metronomes' rod. For our experiments we have used the 
standard frequencies marked on the metronome. These frequencies are given in $BPM$ units.

Before discussing the experimental results in detail, we have to emphasize that, independently of the chosen initial condition, only in-phase synchronization of the metronomes was observed. The reasons for this will be given in a separate section (Section  \ref{sync}). 
 
In the very first experiments we were studying how the chosen frequency influences the detected synchronization level. 
We fixed all the metronomes' frequencies on the same 
marked $\omega_0$ value and placed them symmetrically on the perimeter of the rotating platform as indicated in
Figure \ref{fig3}a. In reality, of course, this does not mean that their frequencies were exactly the same since no two macroscopic physical systems can be exactly identical.  
We initialized the system by starting the metronomes randomly, and let the system composed of the metronomes and platform evolve freely. 

\begin{figure}[ht!]
  \centering
   \resizebox{0.40\textwidth}{!}{%
   \includegraphics{FIG4a.eps}
}
   \resizebox{0.40\textwidth}{!}{%
   \includegraphics{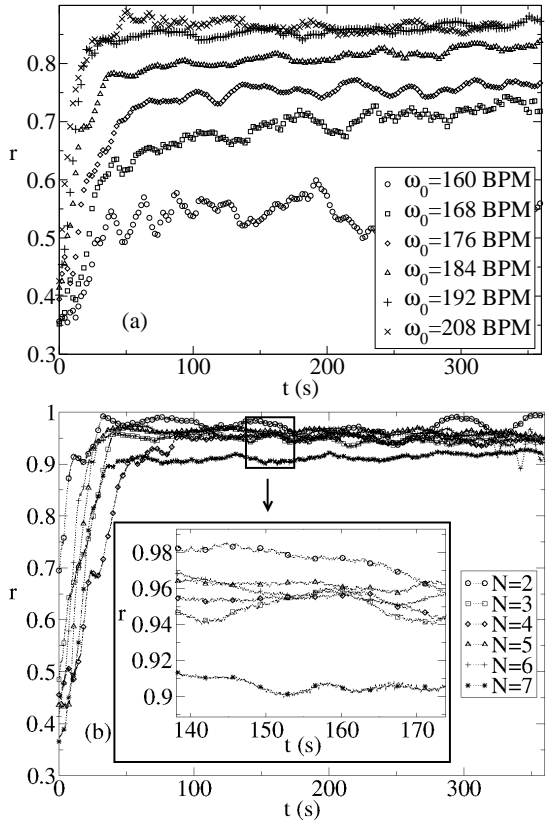}
}

\caption{Dynamics of the order parameter for different natural frequencies and numbers of metronomes. 
(top) Results for seven metronomes, different curves corresponding to different frequencies as indicated in the legends. (bottom) Results are for the fixed frequency ($\omega_0=192\ BPM$) and different numbers of metronomes as
indicated in the legend. On both graphs, the results are averaged across $10$ independent measurements. }
\label{fig5}
 % rng.png: 640x400 pixel, 72dpi, 11x7 cm, bb=0 0 640 400
\end{figure}

For each considered frequency value we made $10$ measurements, collecting data for 10 minutes.  
The dynamics of the computed Kuramoto order parameter averaged across the 10 independent experiments are presented 
in Figure \ref{fig5}a. 

The results suggest that the obtained degree of synchronization increases as the metronomes' natural frequencies increase. The standard deviations of the natural frequencies of the independent 
oscillators are indicated in Table \ref{table2}.
\begin{table}
\centering
\begin{tabular}{|l|l|l|l|l|l|l|}\hline
$\omega_0 (BPM)$  & 160 & 168 & 176 & 184 & 192 & 208\\ \hline
$\sigma$ ($BPM \cdot 10^{-7}$)   &8.4 &7.9 &7.8 &9.8& 8.5 &8.7\\ \hline
\end{tabular}
\caption{Standard deviation of the seven metronomes used for different nominal frequencies.}
\label{table2}
\end{table}

 Since there is no clear trend in this data as a function of $\omega_0$, the obtained result suggests that the observed effect is not due to a
 decreasing trend in the metronomes'  standard deviation.  We have also found that, for the standard metronome frequencies below $160$ BPM, the system 
did not synchronize. It is interesting to note, however, that if one inspects visually or 
auditorily the system, one would observe no synchronization for frequencies already below  $184$ BPM. This means that we are not suited to detect partial synchronization with an order parameter below $r=0.75$. 

In a second experiment we were investigating the influence of the number of metronomes on the synchronization
level. In order to study this, we fixed the metronomes at the same frequency ($\omega_0=192$ BPM) and 
repeated the previous experiment with increasing numbers of metronomes placed on the rotating platform. 

Again, we performed 10 measurements for each configuration so as to obtain accurate results and averaged the observed order parameter. The averaged results are presented in Figure \ref{fig5}b.

Although the standard deviation of the metronomes' natural frequencies (Table \ref{table1}) does not present a clear trend as a function of the number of metronomes, $N$, 
we see a clear trend in the detected synchronization level: increasing the number of metronomes will result in 
a decrease in the synchronization level. 
\begin{table}
\centering
\begin{tabular}{|l|l|l|l|l|l|l|}\hline
N  & 2 & 3 & 4 & 5 & 6 & 7 \\ \hline
$\sigma $  ($BPM \cdot 10^{-7}$)   &5.1 &8.1 &7.5 &7.1 &6.7 &8.5\\ \hline
\end{tabular}
\caption{Standard deviation of the metronomes' natural frequencies for different numbers of metronomes on the rotating platform ($\omega_0=192$ BPM) }
\label{table1}
\end{table}

\section{Theoretical model}
\label{theo}
\begin{figure}[ht!]
  \centering
 \resizebox{0.40\textwidth}{!}{%
   \includegraphics{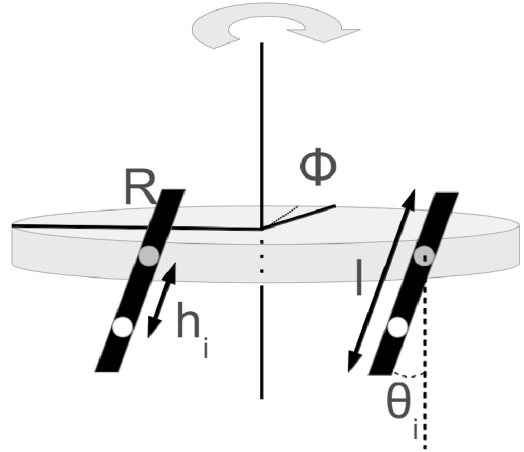}
}
\caption{Schematic view and notations for the considered mechanical model. The white dots denote the center of mass
of the physical pendulums and the gray dots are the suspension axes.}
\label{Fig7}
\end{figure}

Inspired by the model described in \cite{kapit}, it is possible to consider a simple mechanical model for the 
system investigated here.  The model is composed of a rotating platform and physical 
pendulums attached to its perimeter, as is sketched in Figure \ref{Fig7}.

The Lagrange function of such a system is written as:
 \begin{eqnarray}\nonumber
  L&=&\frac{J}{2}\dot \phi^2 + \sum_{i=1}^N \frac{J_i \omega_i^2}{2}+\sum_{i=1}^N \frac{m_i}{2} \Big \{ \Big [ \frac{d}{dt}\Big(x_i+h_i \sin \theta_i\Big)\Big]^2 + \\
 && +\Big[\frac{d}{dt}\Big(h_i \cos \theta_i\Big)\Big]^2\Big\} - \sum_{i=1}^N m_i g h_i(1-\cos\theta_i)
 \label{lagrange}
 \end{eqnarray}
 The first term is the kinetic energy of the platform, the second is the kinetic energy due to the rotation of the pendulum around its center of mass, the third one is the kinetic energy of the pendulum's center of mass, and the last term is the potential energy of the pendulum. In the Lagrangian we have used the following notations:
the index $i$ denotes the pendulums, $J$ is the moment of inertia of the platform with the metronomes on it -- taken
relative to  the vertical rotation axes, $\phi$ is the angular displacement of the platform, $m_i$ is the total mass of the pendulum ($m_i \approx W_1^{(i)}+W_2^{(i)}$, neglecting the mass of the rod), $h_i$ is the distance between the center of mass and the suspension point of the pendulum, $x_i$ is the horizontal displacement of the center of mass of the pendulums due to the rotation of the platform, $\theta_i$ is the 
displacement of the $i$-th pendulum's center of mass, in radians, $J_i$ is the moment of inertia of the pendulum relative to its center of mass and $\omega_i$ is the angular velocity of the rotation of the pendulum relative to its 
center of mass. It is easy to see that $x_i=R\dot \phi$ and $\omega_i=\dot \theta_i$. Assuming now that
the mass of all the weights suspended on the metronomes' bobs are the same ($W_1^{(i)}=w_1$, $W_2^{(i)}=w_2$, and
consequently $m_i=m$), and disregarding the $m_igh_i$ constant terms, one obtains:
\begin{eqnarray}\nonumber
 L'&=&\Big(\frac{J+N m R^2}{2}\Big)\dot \phi^2 + \sum_i \Big(\frac{m h_i^2}{2} + \frac{J_i}{2}\Big)\dot \theta_i^2+\\
&&  + m R \dot \phi \sum_i h_i \cos\theta_i \cdot \dot \theta_i + mg\sum_ih_i \cos\theta_i 
\end{eqnarray}
The Euler-Lagrange equations of motion yield:
\begin{equation}
 (J+NmR^2)\ddot \phi + mR\sum_ih_i[\ddot \theta_i \cos\theta_i-\dot \theta_i^2 \sin\theta_i] = 0
 \label{eqm-ndf1} \\
 \end{equation}
 \begin{equation}
 [m h_i^2+J_i] \ddot \theta_i + mR \ddot \phi  h_i \cos \theta_i+mgh_i \sin\theta_i = 0.
\label{eqm-ndf2}
\end{equation}

The above equations of motion are for a Hamiltonian system where there is no damping (no friction)
and no driving (excitation).  Friction and excitation from the met\-ro\-no\-mes' driving mechanism 
has to be taken into account with some extra terms. The system of equations of motion may be written as
\begin{eqnarray}\nonumber
  (J+NmR^2)\ddot \phi &+& mR\sum_ih_i[\ddot \theta_i \cos\theta_i-\dot \theta_i^2 \sin\theta_i] +\\
&&  + c_{\phi} \dot \phi + \sum_i \mathbb{M}_i= 0 
 \label{metron2a}
\end{eqnarray}
\begin{eqnarray}\nonumber
 [mh_i^2+J_i]\ddot \theta_i &+& mR\ddot \phi  h_i \cos \theta_i+\\
&& +mgh_i \sin\theta_i + c_{\theta} \dot \theta_i= \mathbb{M}_i.
\label{metron2}
\end{eqnarray}
where $c_{\phi}$ and $c_{\theta}$ are coefficients characterizing the friction in the rotation of the platform 
and pendulums, and $\mathbb{M}_i$ are some instantaneous excitation terms defined as
\begin{equation}
\mathbb{M}_i=M \delta(\theta_i) \dot\theta_i,
\label{excitation}
\end{equation}
where $\delta$ denotes the Dirac function and $M$ is a fixed parameter characterizing the driving mechanism of the metronomes.  The choice of the form for $\mathbb{M}_i$ in Equation (\ref{excitation}) means that excitations are given only when the met\-ro\-no\-me's bob passes the $\theta=0$ position.  The term 
$\dot\theta$ is needed in order to ensure a constant momentum input, independently of the metronomes' amplitude. 
It also ensures that the excitation is given in the correct direction  (the direction of motion). 
It is easy to see that the total momentum transferred, $\mathbb{M}_{trans}$, to the metronomes
in a half period ($T/2$)  is always $M$:
\[
\mathbb{M}_{trans}=\int_{t}^{t+T/2} M \delta({\theta_i}) \dot\theta_i dt=\int_{-\theta_{max}}^{\theta_{max}} M \delta({\theta_i}) d\theta_i = M
\]
This driving will be implemented in the numerical solution as
\[
  \mathbb{M}_i = \left\{
  \begin{array}{l l l}
    M/dt & \quad \text{if $ \theta_i{(t-dt)} < 0$ and $ \theta_i{(t)} > 0$ }\\
    -M/dt & \quad \text{if $ \theta_i{(t-dt)} > 0$ and $ \theta_i{(t)} < 0$}\\
    0  & \quad \text{in any other case}\\
  \end{array} \right.
\]
where $dt$ is the time-step in the numerical integration of the equations of motion. Clearly, this driving leads to the same
total momentum transfer $M$ as the one defined by Equation (\ref{excitation}).

The coupled system of equations 
(\ref{metron2a},\ref{metron2}) can be written in a form more suitable for numerical integration: 

\begin{equation}
\ddot \phi =\frac{ mR
\sum_ih_i\dot \theta_i^2 \sin\theta_i - c_{\phi} \dot \phi - \sum_i \mathbb{M}_i +A + B - C}{ D },
\label{eqm1}
\end{equation}

\begin{equation}
 \ddot \theta_i =\frac{ \mathbb{M}_i - mR\ddot \phi  h_i \cos \theta_i - mgh_i \sin\theta_i - c_{\theta} \dot \theta_i}{mh_i^2+J_i}
\label{eqm2}
\end{equation}
where
\begin{align} \nonumber 
 A &= m^2gR\sum_i\frac{h^2_i \sin\theta_i \cos\theta_i}{mh_i^2+J_i}, \\ \nonumber
B &= mRc_{\theta} \sum_i\frac{h_i \dot \theta_i \cos \theta_i}{mh_i^2+J_i}, \\ \nonumber 
C &= mR\sum_i\frac{h_iM_i \cos\theta_i}{mh_i^2+J_i} \text{,} \\ \nonumber
D &= \Big (J+NmR^2-m^2R^2\sum_i\frac{  h^2_i \cos^2 \theta_i}{mh_i^2+J_i}\Big). \nonumber
\end{align}
Now taking into account that the metronomes' bobs have the form sketched in Figure \ref{fig2}b and the 
$L_1$ distances are fixed and assumed to be identical for all the metronomes, 
the $h_i$ and $J_i$ terms of the physical pendulums in our model will be calculated as:

\begin{align}
 h_i=\frac{1}{w_1+w_2} (w_1 L_1-w_2 L_2^{(i)}) \\
 J_i=w_1 L_1^2+w_2 (L_2^{(i)})^2.
\end{align}

\section{Realistic model parameters}
\label{mparam}

The parameters were chosen following the 
experimental device: $w_1 = 0.025$ $kg$, $w_2 = 0.0069$ $kg$, 
$L_1 = 0.0358$ $m$,  $ L_2 \in  [0.019,0.049]m$ depending on the chosen 
natural frequency, $R = 0.27$ $m$ and  $J \in [0.0729,0.25515]kg~m^2$
depending on the number of metronomes placed on the platform. 

The damping and excitation coefficients were estimated as follows. 
For the estimation of $c_{\theta}$, a single metronome on a rigid support was considered. Switching off the excitation mechanism, a quasi-harmonic damped oscillation of the metronome took place. 
The exponential decay of its amplitude as a function of time uniquely defines the damping coefficient, hence a simple fit of the amplitude as a function of  time allowed the determination of $c_{\theta}$. 
Switching the excitation mechanism on lead to a steady-state oscillation regime with constant amplitude. Since $c_{\theta}$ has already been measured, this amplitude value is defined by the excitation coefficient $M$. Solving Equations (\ref{metron2a}) and (\ref{metron2}) for a single metronome and tuning it until the same steady-state amplitude is obtained as in experiments makes the estimation of $M$ possible.  
Now that both $c_{\theta}$ and $M$ are known, the following scenario is considered: all the metronomes are placed on the platform and synchronization is reached. Then the platform has a constant-amplitude oscillatory motion. In order to determine $c_{\phi}$, its value is tuned while solving Equations (\ref{metron2a})-(\ref{metron2}) until the same amplitude of the disk's oscillations is obtained as in the experiments. This way, all the parameters in the model can be related to the experimental quantities. Using the method defined above, we estimated the following 
parameter values: $c_{\theta} = 0.00005$ $kg$ $m^2/s$, $c_{\phi} = 0.00001$ $kg$ $m^2/s$  and $M = 0.0006$ $Nm/s$. 

\section{In-phase synchronization versus anti-phase synchronization of two metronomes}
\label{sync}

As described in the introductory section, many previous works have reported a stable anti-synchronized state in the case of two coupled oscillators \cite{bennet,fradkov,czo}. Due to the  fact that no such stable phase was observed in our experiments (independently of the starting conditions), we feel that investigating this issue is important. Starting from our theoretical model described in Section \ref{theo},
we will show that the in-phase synchronization is favored whenever there are large enough equilibrated damping and driving forces acting on the metronomes. 

\begin{figure}[ht!]
  \centering
  \resizebox{0.40\textwidth}{!}{%
   \includegraphics{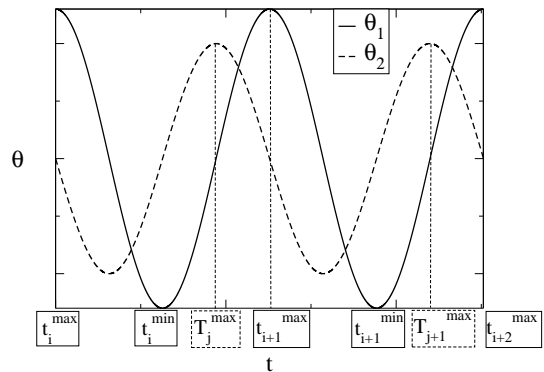}
}
\caption{Dynamics of two metronomes as a function of time, and the quantities used for defining 
the $z$ order parameter.}
\label{fig-opn}
\end{figure}

First, let us investigate the case without any damping and with no driving forces. The equations of motion 
for such a system are given by Equations (\ref{eqm-ndf1}) and (\ref{eqm-ndf2}). Considering the case of two identical 
metronomes ($N=2$) with small-angle deviations ($\theta_{1,2}^{max} <<\pi/2$), we investigate the synchronization 
properties of such a system. The synchronization level will be studied here by an appropriately chosen order parameter for two metronomes, $z$, 
that indicates whether we have in-phase or anti-phase synchronization.
Although we could have used the Kuramoto order parameter for this purpose, 
we decided to introduce a new, more suitable order parameter. Note that
this new order parameter is only useful for small ensembles, because its calculation would be very time consuming for large systems.
In order to introduce a proper order parameter, let us consider the dynamics of two metronomes as a function of time by plotting $\theta_{1,2}(t)$ (Figure \ref{fig-opn}). 
Let us denote the time-moments where metronome $i$ reaches a local minimum and maximum $\theta_1$ values by $t_i^{min}$ and $t_i^{max}$, respectively.
We denote the time-moments where metronome 2 has local maximum $\theta_2$ values by $T_j^{max}$. With these notations, we define two time-like quantities that characterize the average time-interval of the maximum 
position of $\theta_2(t)$ relative to the maximum and minimum positions of $\theta_1(t)$, respectively:

\begin{eqnarray}
t_1=\langle \min_{\{i\}}\{|t_i^{max}-T_j^{max}|\}\rangle_j \\
t_2=\langle \min_{\{i\}}\{|t_i^{min}-T_j^{max}|\}\rangle_j 
\end{eqnarray}

In the above equations the averages are considered over all $j$ maximum positions of $\theta_2(t)$, and the "$\min$" notation  refers to 
the minimal value of the quantity in the brackets.
Now, the $z$ order parameter is defined as:

\begin{equation}
z=\frac{t_2-t_1}{t_2+t_1}
\end{equation}

It is easy to see that $z$ is bounded between $-1$ and $1$. For totally in-phase synchronized dynamics we have $t_1=0$, leading to $z=1$. For totally anti-phase synchronized dynamics $t_2=0$, and we get $z=-1$. Negative $z$ values suggest a dynamic where the anti-phase synchronized states
 are dominant, positive $z$ values suggest a dynamic with more pronounced in-phase synchronized states. 

The $z$ order parameter was estimated numerically for different initial conditions. 
A velocity Verlet-type algorithm was used, and simulations were performed up to a $t=4000$ $s$ time interval, with a  $dt=0.01$ $s$ time-step.

\begin{figure}[ht!]
  \centering
  \resizebox{0.40\textwidth}{!}{%
   \includegraphics{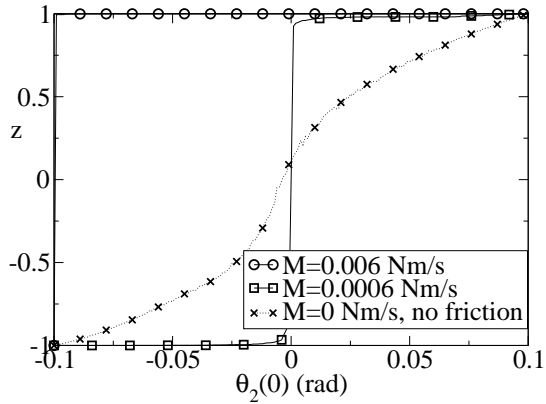}
}

\caption{The $z$ order parameter as function of the initial phase $\theta_2(0)$. 
The results obtained without damping ($c_{\theta}=0$, $c_{\phi}=0$) and driving ($M=0$)
indicate that phase-locking and complete synchronization is possible only if the system starts from
such a situation. For small damping and driving values ($M=6 \times 10^{-5}$ Nm/s), both in-phase and 
anti-phase synchronization is possible. For large driving intensities ($M=6 \times 10^{-4}$ Nm/s), only the in-phase
synchronization is stable. The later is the realistic parameter for the experimental metronome setup.}
\label{q-op}
 % rng.png: 640x400 pixel, 72dpi, 11x7 cm, bb=0 0 640 400
\end{figure}

Initially the deviation angle of the first metronome was chosen as $\theta_1(0)=\theta_{max}=0.1$ rad and $\theta_2(0)$ was chosen in the interval $[-0.1,0.1]$ rad, leading to various initial phase-differences between them. 
The
computed $z$ values as a function of $\theta_2(0)$ are plotted in Figure \ref{q-op}. % elso ketto x-y offset, 3 ik meret, negyedik feher csik merete felul...

The  above results suggest that for the friction-free and undriven case (Figure \ref{q-op}),
synchronization and phase-locking of coupled identical metronomes are possible only if they start either in completely in-phase or completely anti-phase configurations. 
Depending on how the phases are initialized, the ticking dynamics statistically resemble either the in-phase or the anti-phase states, but no phase-locking or synchronization is observable.  
Starting from an arbitrary initial condition a complete in-phase or anti-phase synchronization is possible only if there is dissipation and driving. For small dissipation and driving values
both the in-phase and anti-phase synchronization are possible, as the results obtained for $M=6 \times 10^{-5}$ Nm/s suggests. In this limit in-phase synchronization will
emerge if the initial phases are closer to such a situation. Alternatively, if the initial conditions resemble the anti-phase configuration, a stable anti-phase synchronization emerges.   
For higher dissipation and driving values (characteristic for our experimental setup, $M=6 \times 10^{-4}$ Nm/s ) this apparently symmetric picture breaks down, and the in-phase synchronization
is the one that is stable. Anti-phase synchronization is unlikely to be observed; it will appear only in the case when the two metronomes  are started exactly in anti-phases ($\theta_2(0)=-\theta_1(0)$). 

In view of these results, one can understand
why only the stable in-phase synchronized dynamics was observed in our experiments.
The results also emphasize the importance of using realistic model 
parameters in order to reproduce the observed dynamics. 

\section{Simulation results for several metronomes}
 
\begin{figure}[ht!]
  \centering
  \resizebox{0.40\textwidth}{!}{%
   \includegraphics{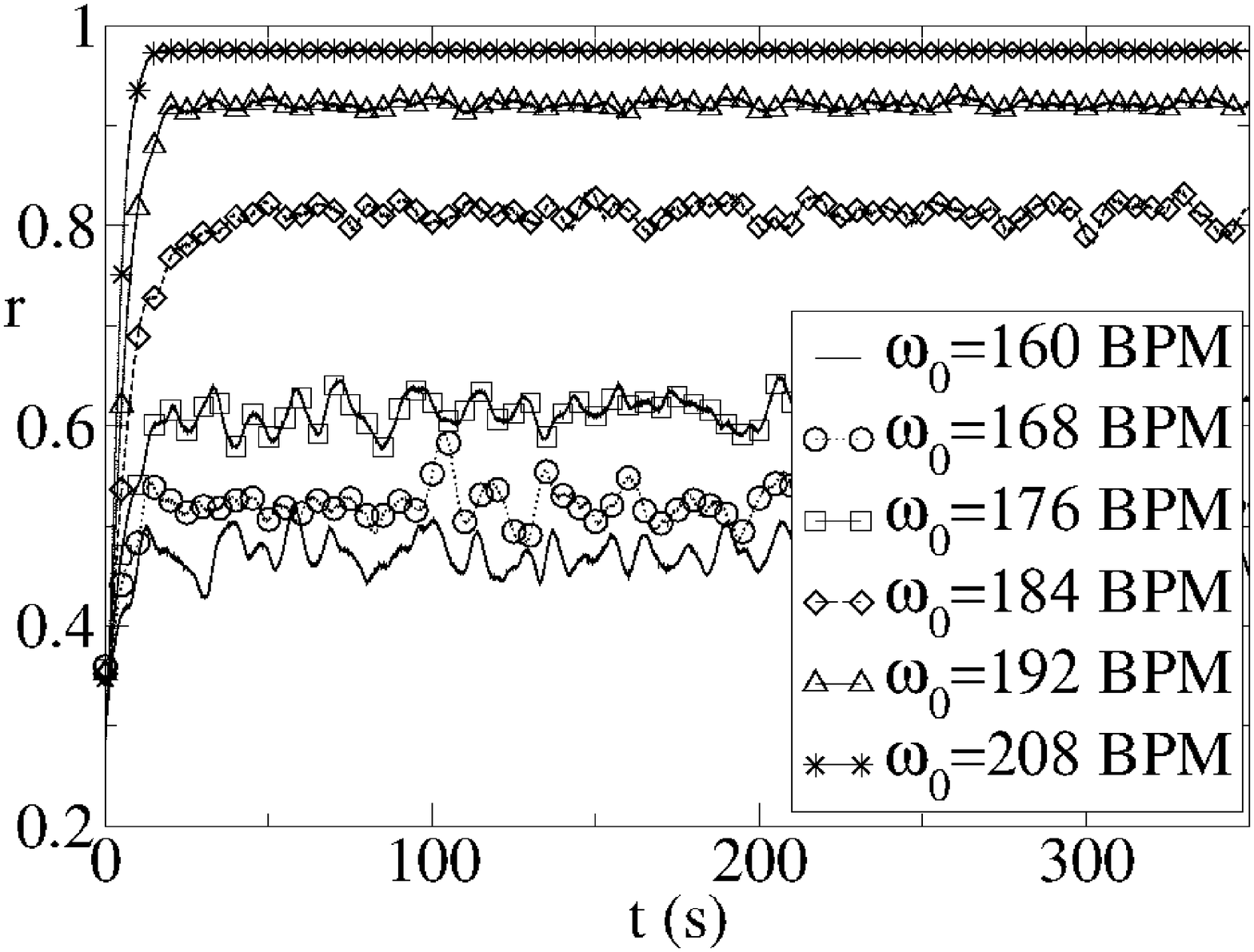}
}
  \resizebox{0.40\textwidth}{!}{%
   \includegraphics{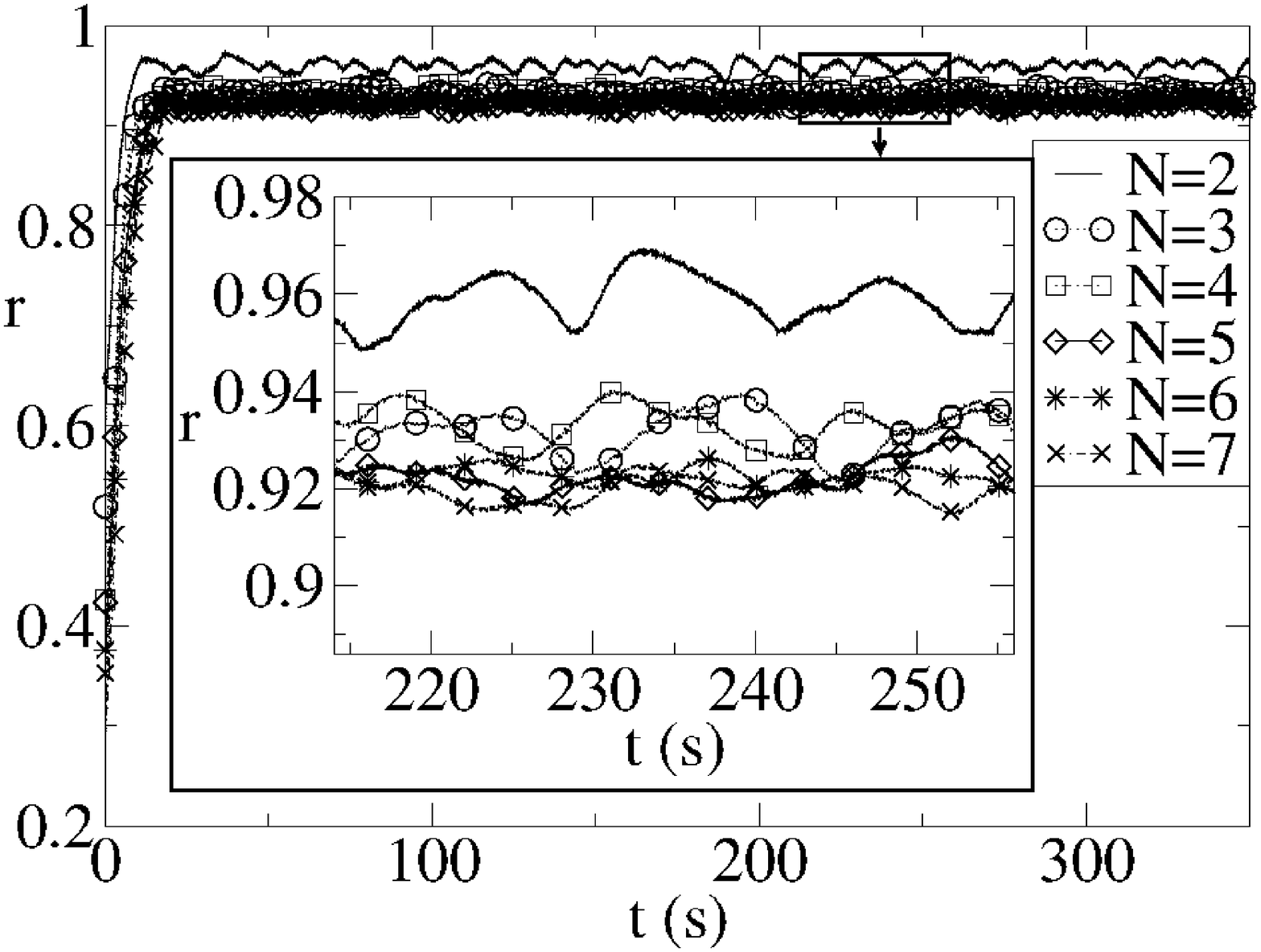}
}

 \caption{Simulation results for the time-evolution of the Kuramoto order-parameter. 
(top) Results for the same natural frequencies as the ones used in the experiments. 
(bottom) Results for the same number of pendulums, and nominal beat frequency ($\omega_0=192$ BPM) as the ones used 
in the experiments. For both graphs, the presented results are an average over $100$ independent simulations. The corresponding 
experimental results are presented in Figure \ref{fig5}. }
\label{fig8}
\end{figure}

Using the model defined in Section 4, our aim here is to theoretically understand the experimentally obtained trends.
The equations of motion (\ref{eqm1}),(\ref{eqm2}) were 
numerically integrated using a velocity Verlet-type algorithm as the integration method. A time-step 
of  $dt=0.01$ $s$ was chosen.  First we intended 
to explain the experimental results presented in Figure \ref{fig5}. 
Seven metronomes with the same $\omega_i$ natural frequencies as the experimentally measured ones were considered,
and the time-evolution of the Kuramoto order parameter was computed. Results obtained for different $\omega_0$ frequency values 
are presented in the top panel of Figure \ref{fig8}.  For the sake of better statistics we averaged the results of $100$ simulations.

The obtained results are in good agreement with the experimental results presented
in Figure \ref{fig5}.  

Following our experiments, we have also studied the time-evolution of the order parameter 
for different numbers of pendulums, setting the same $\omega_0=192$ BPM natural frequency as in the 
experiments. Again, we averaged the results for $100$ 
independent simulations. The obtained trend is sketched on the bottom panel of Figure \ref{fig8}.

The trend of the simulation results is in agreement with the experimental ones: increasing the number of metronomes
results in a decrease in the observed synchronization level. In simulations, however, this decrease is not as
evident as in the experiments.  The reason for this could be the oversimplified manner in which we have handled the differences between the metronomes. In our model, the only difference between the metronomes are in the $L_2^{(i)}$ values (the
distance of the movable weight from the  horizontal suspension axes, see Figure \ref{fig2}). In our simulations, the non-zero spread of these values is the sole source of the $\sigma$ standard deviation for the frequencies $\omega_i$. However, in reality many other parameters of the metronomes  are different, leading to more different model parameters in their equations of motion. As  a result of this, a more pronounced variation in the synchronization level is expectable.

\begin{figure}[ht!]
  \centering
 \resizebox{0.40\textwidth}{!}{%
   \includegraphics{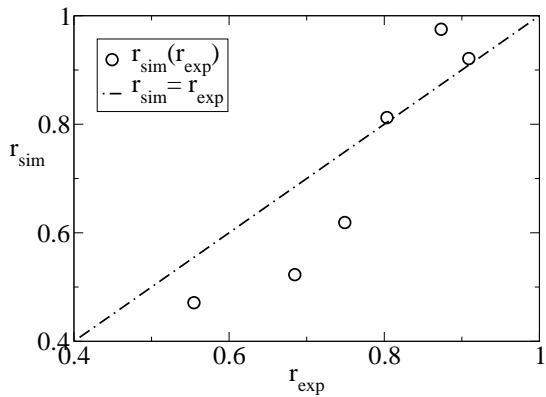}
 }

\caption{Comparison between the simulated and experimentally obtained equilibrium synchronization level. Circles represent the results obtained for the 
$\omega_0$ frequencies shown in Table 1. The straight dashed line indicates the optimal $r_{sim}=r_{exp}$ limit.}

\label{comp}
 % rng.png: 640x400 pixel, 72dpi, 11x7 cm, bb=0 0 640 400
\end{figure}

In spite of the above discussed discrepancy, the simulation results suggest that our 
model with realistic model parameters works well for describing the 
dynamics of the coupled metronome system. In order to
illustrate the effectiveness of our approach more quantitatively, we have plotted the simulated equilibrium synchronization level, $r_{sim}$, as a function of the experimentally determined value, $r_{exp}$, for the case of $N=7$ metronomes.
 The plot from Figure \ref{comp} suggest that there is a satisfactory correlation.

 Thus, one can investigate several interesting cases through simulations that are not feasible experimentally. 
Many interesting questions can be formulated this way. Here we focus however only on clarifying the problems that we have investigated experimentally, 
namely the influence of the number of oscillators and the chosen natural
frequency on the observed synchronization level. 

Computer simulations will allow us to consider a higher number of metronomes and
will also allow for a continuous variation of the metronomes' natural frequencies. Particularly, we are interested in clarifying whether, in the thermodynamic limit ($N\rightarrow \infty$), there
is a clear $\omega_0=\omega_c$ frequency threshold below which there is no synchronization in a system
with fixed standard deviation ($\sigma$) of the metronomes' frequencies.  Also, we would like to show that the reason for not obtaining 
a complete synchronization ($r=1$) of the metronomes is the finite value of $\sigma$.

\begin{figure}[ht!]
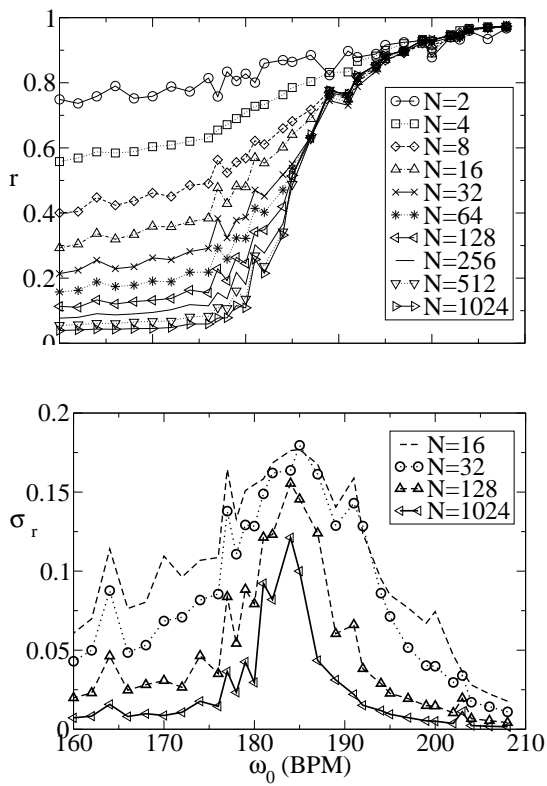

  \centering
 \resizebox{0.40\textwidth}{!}{%
   \includegraphics{FIG10a.eps}
}
 \resizebox{0.40\textwidth}{!}{%
   \includegraphics{FIG10b.eps}
}

\caption{Simulation results for the Kuramoto order parameter (top) and its standard deviation, $\sigma_r$, for the $100$ computational
experiments (bottom) 
as a function of the $\omega_0$ frequency.  Different curves are for different numbers of metronomes as indicated 
in the legends. In each case, the standard deviation of the metronomes' natural frequency $\omega_i$  is fixed as close as possible to $\sigma=8 \times 10^{-7}$ BPM.}

\label{fre}
 % rng.png: 640x400 pixel, 72dpi, 11x7 cm, bb=0 0 640 400
\end{figure}

Considering a normal distribution of metronomes' natural frequency $\omega_i$ with a fixed standard deviation 
around the mean value of the standard deviations presented in Table 1 ($\sigma=8 \times 10^{-7}$ BPM), we first studied how the Kuramoto order parameter, $r$, varies 
 as a function of $\omega_0$. Results obtained for a wide range of the number of metronomes, $N$,
are plotted in the top panel of Figure \ref{fre}. 

The results plotted in Figure \ref{fre} suggest that, in the $N\rightarrow \infty$ limit, a clear phase-transition like phenomenom emerges. Around the value of $\omega_c=185$ BPM the order parameter
exhibits a sharp variation, which becomes sharper and sharper as the number of metronomes is increased. This is a clear sign of phase-transition like behavior. 
Plotting the standard deviation of the order parameter values obtained from different experiments, we get a characteristic peak around the $\omega_c=185$ BPM value. As is expected for a phase-transition-like phenomenon this peak narrows as the number of metronomes on the disk increases.

\begin{figure}[ht!]
  \centering
  \resizebox{0.40\textwidth}{!}{%
   \includegraphics{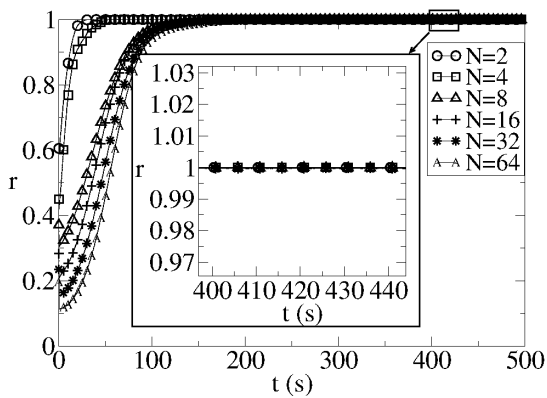}
}   
 \caption{Simulation results for identical metronomes. The time evolution of the order parameter for various numbers of metronomes, ranging from 2 to 64, is plotted. ($\omega_0=192$ BPM)}
\label{fig11}
 % rng.png: 640x400 pixel, 72dpi, 11x7 cm, bb=0 0 640 400
\end{figure}

Our next aim is to prove that the reason for not reaching the $r=1$ complete synchronization is the 
finite standard deviation of the metronomes' natural frequencies.  
Simulations with up to $64$ identical metronomes with $\omega_0=192$ BPM were considered, 
and the $r(t)$ dynamics of the Kuramoto order parameter was investigated. Results for different numbers of metronomes
are plotted in Figure \ref{fig11}.  From these graphs one can readily observe that in each case 
the completely synchronized state emerged. This proves that the lack of complete synchronization is 
due to the finite spread in the metronomes' natural frequencies. From this simulation we have also learned that variations of the 
equilibrium order parameter value as a function of $N$ is also due to the finite $\sigma$ value. 

\section{Conclusions}

The dynamics of a system composed of coupled metronomes was investigated both 
by simple experiments and computer simulations. We were interested in finding the 
conditions for the emergence of synchronization.  Contrarily to many previous studies, here the
problem was analyzed not from the viewpoint of dynamical systems, but from the viewpoint of collective behavior
and emerging synchronization. 

The experiments suggest that there is a limiting 
natural frequency of the metronomes below which spontaneous synchronization is not possible. 
By increasing the frequency above this limit, partial synchronization will e\-mer\-ge. The obtained synchronization level increases monotonically as the natural frequency of the oscillators increases. 
The experiments also suggest that increasing the number of metronomes in the system leads to a decrease of the observed synchronization level. 

In order to better understand the dynamics of the system a realistic model was built. We have shown that damping due to friction forces and the 
presence of driving are both important in order to understand the emerging synchronization. 
The parameters of the model were fixed in agreement with the experimental conditions and the equations of motion were integrated numerically. The model proved to be successful in describing the experimental results, and reproduced the experimentally observed 
trends. The model allowed a fine verification of our findings regarding the conditions under which spontaneous synchronization emerges and the trends
in the observed 
synchronization level.  Computer simulations suggested that, 
for an ensemble of metronomes with a fixed standard deviation of their natural frequencies, the order parameter increases as a function of the
metronomes' average frequency, $\omega_0$. The model also suggests that this increase happens sharply for large ensembles, closely resembling a phase-transition like phenomenon. 
With the help of the simulations we have also shown that the reason behind an incomplete synchronization ($r=1$) is the finite spread of the
metronomes' natural frequencies ($\sigma \ne 0$).

The successes of the discussed model opens the way for many further studies regarding the dynamics of this
simple system. Indeed, many other interesting questions can be  formulated regarding the influence of the met\-ro\-nome and 
rotating platform parameters on the obtained synchronization level and the observed trends. Also, one can study  
systems where the metronomes or groups of met\-ro\-no\-mes  are fixed to different natural frequencies, or where
there is an external driving force acting on the system. The discussed model has the advantage that the equations of motion are easily 
integrable and the model parameters are realistic, with a direct connection to the parameters of an experimentally realizable system.

Finally, we hope that the novel experimental setup and the results presented here will help in clarifying 
some aspects for one of the oldest problems in physics, namely the spontaneous synchronization of coupled pendulum clocks.  Although several similar problems 
have been considered in previous studies, we have shown that there are still many fascinating aspects that one can investigate in this simple mechanical system. 

\nocite{*}

\section{Acknowledgments}
Work supported by the Romanian IDEAS research grant  PN-II-ID-PCE-2011-3-0348. The work of 
B.Sz. is supported by the POSDRU/107/1.5/S/76841 PhD fellowship. 
B. Ty. acknowledges the support of Collegium Talentum Hungary. We thank E. Kaptalan from 
SC Rulmenti Suedia for constructing the rotating platform.

%
% BibTeX users please use
% \bibliographystyle{}
% \bibliography{}
%
% Non-BibTeX users please use

\end{document}